\documentclass[preprint,12pt]{elsarticle}

\usepackage{amssymb}
\usepackage{amsmath,amssymb,amsfonts}
\usepackage{algorithmic}
\usepackage{graphicx}
\usepackage{textcomp}
\usepackage{xcolor}
\usepackage{url}
\usepackage{float}
\usepackage{amsmath}
\usepackage{subcaption}  
\usepackage{subfig}
\usepackage{booktabs}
\journal{Arxiv }

\begin{document}

\begin{frontmatter}



\title{ Enhancing Kubernetes Resilience through Anomaly Detection and Prediction}


\author[uoa]{V. Anemogiannis}
\ead{bgane@di.uoa.gr}
\author[uoa]{B. Andreou}
\ead{bcand@di.uoa.gr}

\author[uoa]{K. Myrtollari}
\ead{kostasmyrto@di.uoa.gr}

\author[uoa]{K. Panagidi}
\ead{kakiap@di.uoa.gr}

\author[uoa]{S.~Hadjiefthymiades%%\fnref{fn1,fn3}		
}
\ead{shadj@di.uoa.gr}

\address[uoa]{Department of Informatics and Telecommunications,National and Kapodistrian University of Athens,Greece }

%% Abstract
\begin{abstract}
Kubernetes, in recent years, has become widely used for the deployment and management of software projects on cloud infrastructure. Due to the execution of these applications across numerous Nodes, each one with its unique specifications, it has become a challenge to identify problems and ensure the smooth operation of the application. Effective supervision of the cluster remains a challenging and resource intensive task. This research work focuses on providing a novel framework system maintainer in order to overview all the possible resources in Kubernetes and pay the attention to specific parts of the cluster that may be showcasing problematic behavior. The novelty of this component rises from the use of cluster graphical representation where features, e.g. graph edges and neighboring nodes, are used for anomaly detection. The proposed framework 
defines the normality in the dynamic enviroment of Kubernetes and the output feeds the supervised models for abnormaliry detection presented in user-friendly graph interface. A variety of model combinations are evaluated and tested in real-life environment.
\end{abstract}


%% Keywords
\end{frontmatter}


\section{Introduction}

One of the most important developments in recent years of software and application development is the switch to cloud computing \cite{Gulhane:2022}. Instead of designing applications as a single monolithic entity, designed to run on a local machine, developers have switched their focus in creating smaller and independently developed and deployed micro services delivered to the users over the internet \cite{Liu:2020}. One of the key ingredients of this transition is the Container. A Container is a small and light virtual machine that can house a micro service. The combination of micro services, spread across multiple Containers can be used to create an application a user can access. 

In order to manage these Containers, a Container orchestration platform is needed. One of the most popular Container orchestration platforms is Kubernetes \cite{kubernetes} \cite{Shah:2019}. Kubernetes houses the Containers inside Pods. Pods are the smallest unit that can be deployed in Kubernetes. These Pods are spread across various Nodes of different specifications. The collection of these Nodes, used to run an application, constitutes the Cluster. This approach has numerous advantages including scalability and operational resilience. 

Managing the Pods spread across multiple Nodes, that are perhaps geographically separated, can be a challenge. Pods as well as Nodes can present problems in their operations that could jeopardise the smooth operation of the entire application. Monitoring tools exist \cite{Prometheus} for accessing Kubernete's metrics but understanding them and making predictions on how problems in one component can effect others, constitutes a difficult challenge, especially for a system maintainer without extensive experience.
% on where the problems can spread can constitute a difficult challenge, especially for a novas system maintainer without deep domain expertise.

Various schemes have been proposed to address this challenge as described in Subsection \ref{Algorithmic Anomaly and Error Prediction on Kubernetes Resources}. These works leverage machine learning techniques in order to detect anomalous behavior and notify the system maintainer to address it. Most solutions focus, however, on the individual components, ignoring their interconnected nature. A Pod misbehaving inside a Node can have negative effects on other Pods housed inside the same Node. In addition, most works assume a level of domain expertise in the creation of datasets of normal and anomalous behavior to train the models. 

The contribution of this research in this field, is the proposal of a graph representation of the Kubernetes Cluster, fed with monitoring information and updated in real time in the anomaly detection process. A user-friendly interface of graph dynamically produced by unsupervised techniques based on the dynamic resources in the Kubernetes cluster is combined with supervised techniques in order to identify possible anomalies to the system. This will allow the anomaly detection component to easily leverage information from connected Kubernetes components, using the edges of the graph. The proposed solution allows to easily integrate various supervised and unsupervised methods as models. 

This paper is organised as follows: Section II showcases related works to this research. Section III breaks down the proposed solution. The model selection and the experiments performed are detailed in Section IV. Finally, the conclusions and future work are presented in section V.

\section{RELATED WORK}

\subsection{Algorithmic Anomaly and Error Prediction on Kubernetes Resources}
\label{Algorithmic Anomaly and Error Prediction on Kubernetes Resources}

Various systems have been suggested for detecting anomalies on monitoring data. In their work,  Kosińska et al. present the Kubernetes Anomaly Detector (KAD) system, that follows a streamlined workflow from data acquisition to model selection and deployment \cite{Kosińska:2022}. The authors evaluate four models: Seasonal Autoregressive Integrated Moving Average, Hidden Markov Model, Long Short-Term Memory, and Autoencoder.
Their system can be classified as semi supervised due to the assumption that the training data is non anomalous. In addition, KAD offers automatic model selection with criteria of validation loss minimilization. Kosińska et al. assert that having multiple models to choose from provides more flexibility for the type of data and speed requirements for anomaly detection.

Qingfeng et al. introduce an Anomaly Detection System for container based microservices. It is  comprised of monitoring, data processing, and fault injection modules used for the training of the models \cite{Qingfeng:2018}. The monitoring module gathers performance data from containers across clusters, while the data processing module analyzes this data to detect anomalies. The authors claim that their system can identify anomalies such as CPU hogging, memory leaks, packet loss, and network latency using supervised machine learning. The algorithms tested in this work, include Support Vector Machines, Nearest Neighbor Classifier, Naive Bayes, and Random Forest. To train the algorithms, the authors used the fault injection component by writing scripts to simulate various types of faults in their system. Faults simulated include high CPU consumption, memory leaks, network package loss, and network latency increase.

Tien et al. introduce KubAnomaly, a security monitoring system tailored for Kubernetes environments, aimed at detecting anomalous behavior at the container runtime \cite{Tien:2019}. Their system aggregates monitoring logs from containers, extracts features, and inputs them into a 4 layer fully connected linear neural network with activation functions and dropout layers. KubAnomaly labels containers as Normal or Anomalous, on a 10 second interval. The authors evaluated KubAnomaly by creating two datasets of varying complexity. They simulated normal behavior by deploying a service and observing typical usage patterns, while they induced anomalies by generating requests to simulate DDOS attacks, autotraversers attempting to extract information from containers, and hacking attempts aimed to compromise the web service.

These works are rooted at the component performance question, mostly the Container, and perform the anomaly detection using metrics originated by the component in question. This strategy diverges from a holistic view and treatment of the whole cluster and how it could affect the component's performance in the near future. In addition, all these works assume a domain expert in order to create the datasets that train the models.

\subsection{Our Contribution}

In contrast to the aforementioned solutions, this work tries to provide a solution that takes into account the interconnected nature of components inside the Kubernetes cluster. In addition, it tries to learn the rules that create these anomalies without making presuppositions on the data.
Our contribution to the field of Kubernetes Anomaly Detection and Prediction lies in:
\begin{itemize}
    \item implementation of a framework providing a holistic view of how anomalies spread within the Kubernetes cluster.
    \item the use of unsupervised models to dynamically define the "normality" of Kubernetes ecosystem
    \item the use of the supervised models for anomaly detection based on the the trained "normality"
    \item a model toolbox offered to users to create a unique combination of unsupervised and supervised models
    \item Present the results in a user friendly manner and track the source of an anomaly.
\end{itemize}

Ultimately, our goal is to develop an intuitive and user-friendly anomaly detection tool that provides actionable insights into the health and performance of Kubernetes clusters to empower operators with the information needed to proactively address issues and ensure the reliability of their containerized applications.

\section{PROPOSED SOLUTION}

 \subsection{Background - Systems and Monitoring}

The closest to anomaly detection mechanism occurs by Kubernete's own self healing capability. This consists of a reactive mechanism that detects containers that fail or do not pass user defined tests and restarts them or replaces them, in order to ensure and maintain application availability. In addition, Kubernete's Horizontal Pod Autoscaler dynamically adjusts the number of Pods on a given Deployment as a reaction to specified metrics such as CPU utilization. The above mentioned mechanisms, may be fairly effective but are considered reactive in nature. They respond to events as they occur without the ability to predict and prevent anomalies \cite{kubernetes}.

 In order to detect any anomalies  and predict if an undesired behavior takes place, there is a need for access to Kubernetes' metrics. This is where Prometheus comes into place. Prometheus collects numeric data from components inside the Kubernetes Cluster and presents it as a time series data stream. Prometheus not only collects data such as CPU and memory utilization, but also metrics like the number of requests on a given service \cite{Prometheus}. In addition, Thanos complements Prometheus by extending its capabilities. This is done by enabling long-term storage, global querying and cross cluster aggregation of metrics data. Using Thanos, developers can achieve scalable and highly available monitoring solutions by aggregating and querying metrics data from distributed Kubernetes clusters \cite{thanos}.

\begin{figure}
    \centering
    \includegraphics[width=0.9\linewidth]{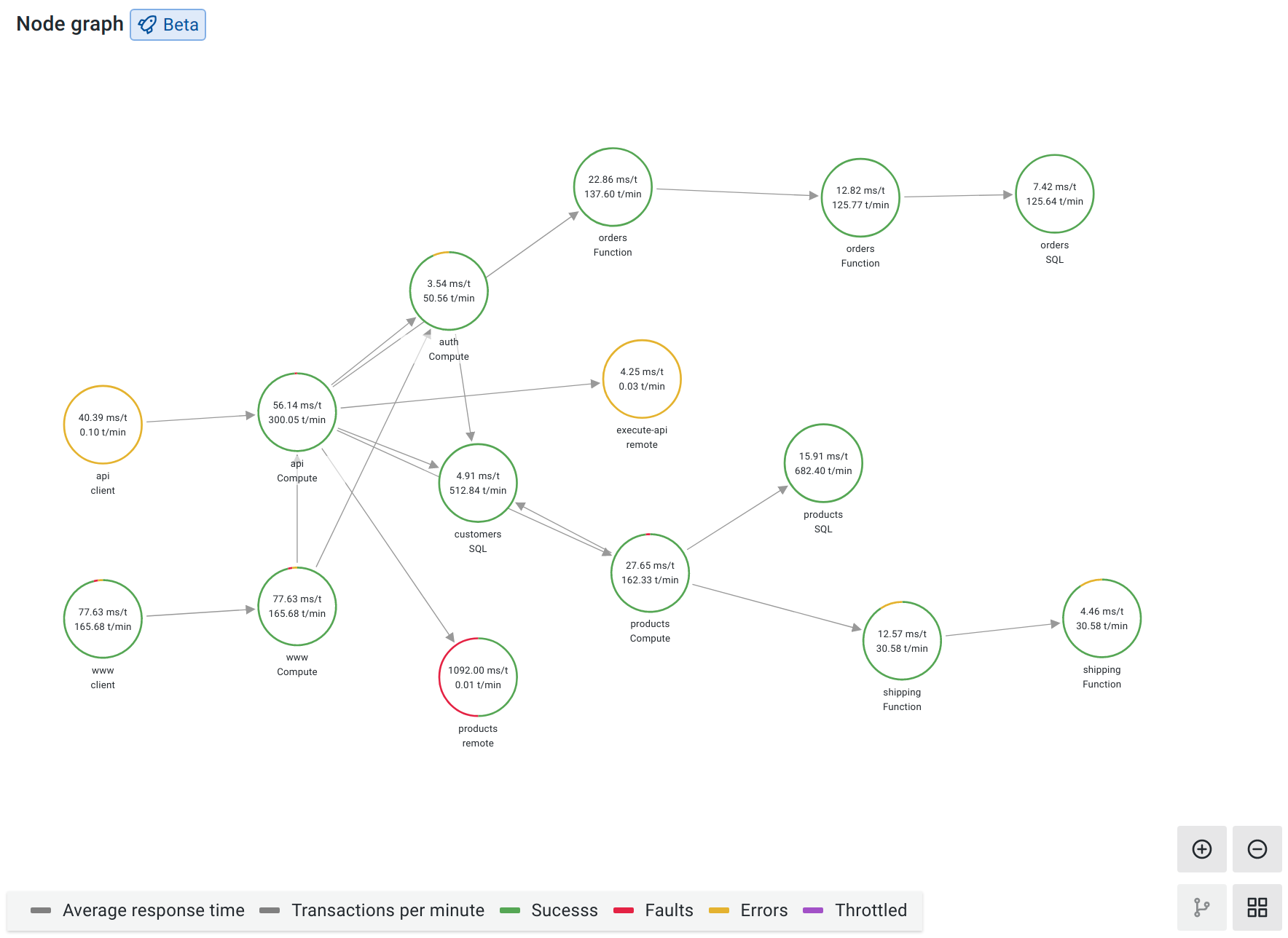}
    \caption{Graphana's Graph Visualization of Monitoring Data}
 \label{fig_graphana}
\end{figure}

 Last, Grafana is an important component in the monitoring ecosystem as it offers powerful query visualization as well as alerting functionalities based on Prometheus's, and by expansion, Thanos's metrics. Notably for this research, Grafana provides a visual Node graph as shown in Fig. \ref{fig_graphana}, where Nodes represent various applications, services, and other components, and Edges symbolize relationships between them, such as requests. Within this graph, Nodes and Edges contain pertinent information about the entities they represent, while also offering the capability to highlight important metrics and health data. Using this visualization tool enables the user to analyze and understand the relationships and dependencies within the kubernetes environment, enabling proactive anomaly detection and efficient troubleshooting by an experienced system maintainer \cite{grafana_nodegraph} as shown in \ref{fig_graphana}.

\subsection{Architecture}
An important aspect of this work is the Database used in order to store the information of the Kubernetes Cluster. The Kubernetes cluster could be viewed as a Dynamic and Heterogeneous Graph, where nodes signify Kubernetes components such as Pods and Deployments, while edges correspond to the connections between the aforementioned components. The Database chosen needs to be able to store this graph, a relationship heavy model, while offering an easy to use query language (Cypher Query Language) and visualization capabilities. 

\subsubsection{ Monitoring Component}
Let consider a Kubernetes cluster with all the available resources. Prometheus has an additional overview of the resources in the cluster. We propose a Monitoring component as shown in Fig. \ref{fig_anom_comp} which consumes the metrics such as CPU saturation and memory utilization,  stores the results from its own anomaly detection process on the Prometheus data inside the corresponding nodes of a graph and ensures that the information stored inside the nodes and edges of the graph is updated using the data coming from Prometheus.

\begin{figure}
    \centering
    \includegraphics[width=1.0\linewidth]{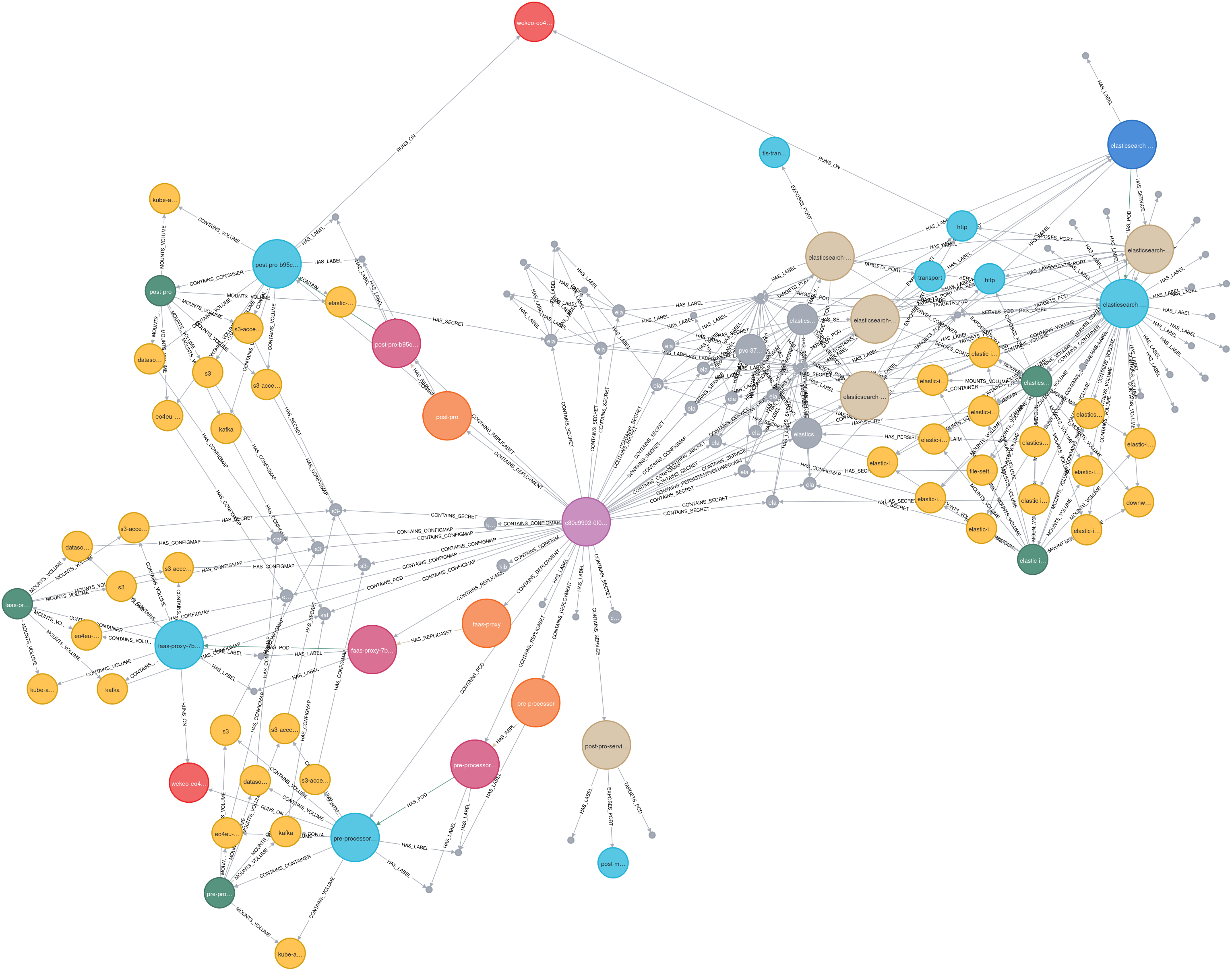}
    \caption{sing Neo4j to Represent a Workflow in the Kubernetes Cluster. The red node at the top symbolizes the Node the Workflow runs on. The pink node at the middle represents the Namespace created for this Workflow. The orange nodes represent Deployments, while the pink ones represent Replica Sets. The big light blue nodes represent Pods while the green nodes represent the Containers. There are also other types of nodes inside the graph such as Services, Stateful Sets, Ports, Labels, etc.}
    \label{fig_workflow}
\end{figure}

Neo4j selected as long as offers the ability to store information inside the nodes and the edges of the graph \cite{Lopez:2015, Rodriguez:2010, Regmi:2021} as the backbone of the architecture.This makes it possible to save the monitoring data as well as anomaly classification results inside the nodes. The user has easy access to that information via the User Interface or by a query, gaining access to a health overview of the entire cluster as shown in Fig. \ref{fig_workflow}.

Utilizing a graph database, the system, can solely rely on the graph for the anomaly detection and prediction task, without the need for direct communication with the Kubernetes or Prometheus APIs. This enables the component to be scalable, flexible and resilient, making it well suited for dynamic and distributed environments that can be represented on a graph database.

\begin{figure}
    \centering
    \includegraphics[width=0.8\linewidth]{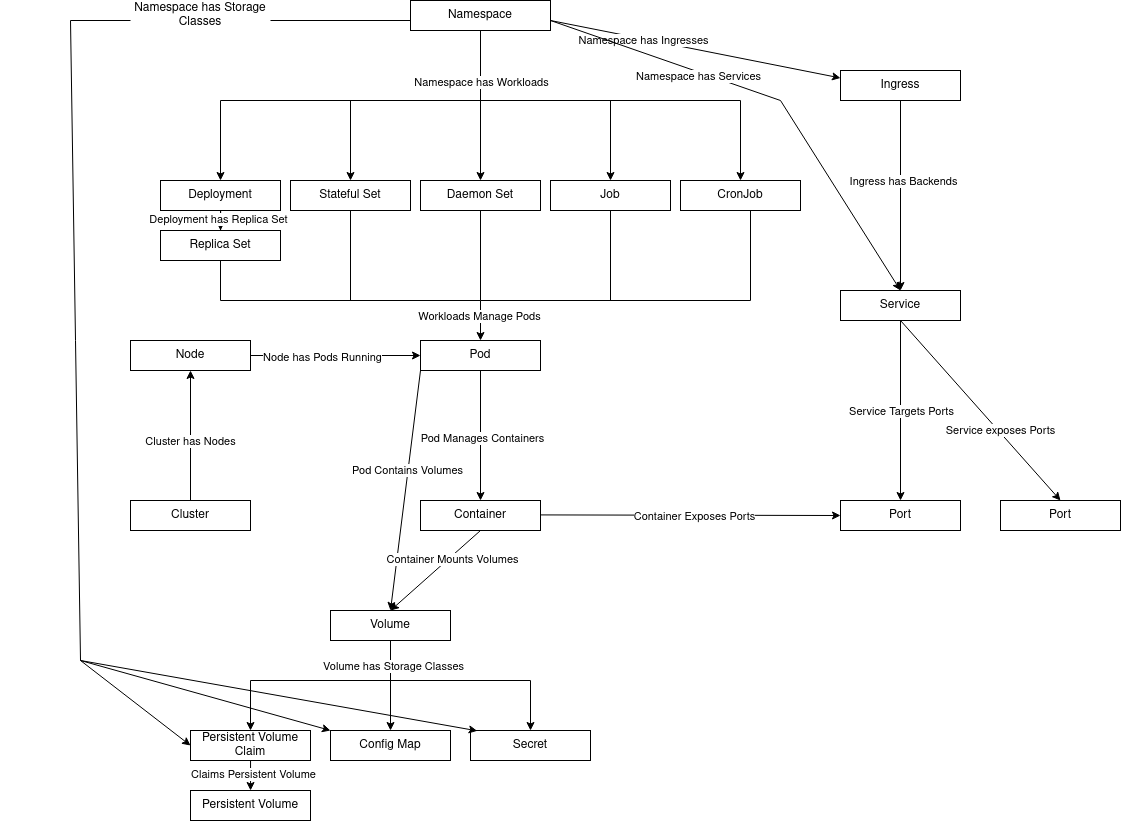 }
    \caption{Kubernetes Resource Model}
    \label{fig_KRM}
\end{figure}

Lastly, the deployment of the Resource Registry component as shown in Fig. \ref{fig_KRM} in the same application layer to the component described in this work, as shown in Fig. \ref{fig_arch}, ensures the query capabilities and access to the real time augmented representation of the cluster, for the prediction process to take place. 

\subsubsection{Anomaly Detection and Prediction on Future Spreading Strategy}

The nature of the described Kubernetes graph, that could be characterized by its heterogeneity, meaning that multiple types of nodes coexist in the graph and represent different kubernetes components, each with its distinct monitoring metrics, and its dynamic nature, meaning nodes and edges are constantly added or removed from the graph following the events in the kubernetes cluster, makes employing conventional machine learning techniques a challenge. The use of the entire graph as input makes it really challenging to expect as output the same graph with annotated anomaly values for each node.

To address the challenge described, the anomaly detection and prediction process is segmented based on the types of the nodes within the graph. This means that a Pod is evaluated against other Pods and so forth, for the different component types. It is worth mentioning that this segmentation does take into account the metrics of the connected components to the one being evaluated using the edges of the graph to reach them. 

Not all components in the cluster have a physical meaning. Many  Kubernetes components serve as an abstraction layer and don't have their own monitoring information. For example, the Deployment functions as an abstraction layer for the Replica Sets that it manages, and the Replica Set in turn, serves as an abstraction layer for the Pods that it manages. In such cases, these components will act as aggregators to help consolidate metrics from related components.

\subsubsection{Dividing the Anomaly Detection and Prediction Process into 2 Parts}

Without domain expertise, a dataset for what behavior constitutes as anomalous for each component cannot be created. This realization leads to the division of the anomaly detection and prediction process into two distinct parts, the unsupervised and the supervised phases.

The anomaly detection and prediction process is divided into two distinct parts, the unsupervised and the supervised phases. This is done due to the lack of domain expertise as to what constitutes anomalous behavior, making the creation of a training dataset a challenge. 

\subsubsection{Unsupervised Part} 

In the unsupervised phase, the question of what is defined as a normality is addressed. The dynamic nature and the easiness to change pods, nodes, memory make the performance overview of a cluster a challenging task. In addition resources shall be treated similarly based on the nature of the resource, e.g. a Pod shall be evaluated against other Pods and so forth, for the different component types.

In unsupervised anomaly classification models are utilized on datasets of observations from the graph. The task consists of the utilization of existing observations to discern, for each component type, which observations stray from the normality. It is worth noting that all observations could be deemed normal by the model, meaning it did not find any outlier observations.

The outlined process leads to the construction of a dataset of normal and anomalous behavior, using the previous observations. As the state of the cluster evolves over time, in order to avoid drifting, it is important to re-execute this part of the process. This is done in order to incorporate in the anomaly comparison the latest observations of the graph. The more observations that are incorporated to the decision process, the less the likelihood of normal behavior to be classified as anomalous will be.

\subsubsection{Supervised Part}

Using the dataset with data labeled as normal or anomalous from the unsupervised part, traditional classification machine learning models can now be used. The goal is to perform two class classification in order to determine for future observations their anomaly status. It is worth mentioning that the models used for this part need to be able to provide a continuous output due to the nature of the task, that requires output to be a probability value. It is worth mentioning that the models used for this part need to have the ability to give probability as output since the anomaly classification is a continuous task. 

The trained models can be used for classification on the periodic scrapings of the graph. The outputs of the models are used to annotate the nodes of the graph with their given anomaly score.

When the unsupervised part is re-executed to take into account the latest observations, the supervised models are also retrained on the new labeling.

\subsubsection{Anomaly Detection and Prediction on Future Spreading Workflow}

\begin{figure}
\centering
\includegraphics[width=0.8\linewidth]{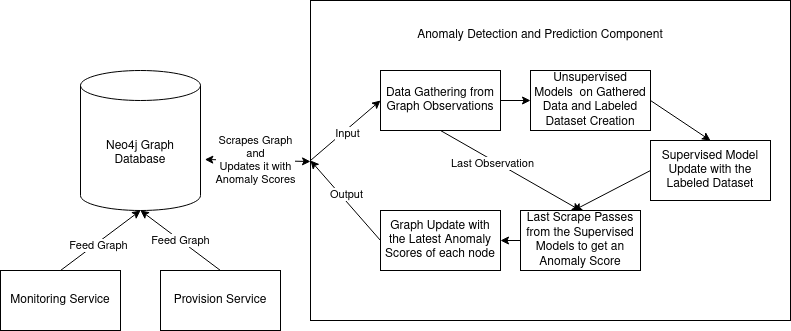}
\caption{Anomaly Detection and Prediction Component Architecture}
\label{fig_anom_comp}
\end{figure}

The Anomaly Detection and Prediction Workflow consists of two functions that are executed at user specified intervals: the "update graph" and the "update models" functions. The first one is invoked to gather a graph observation and label the nodes of the graph with anomaly values, while the later is called in order to retrain the models with the latest observations. A simplified architecture is illustrated in Fig. \ref{fig_anom_comp}.

\paragraph{Update Graph}

When calling the "update graph" function, two tasks take place. The update of the historic graph observations with the new ones and the "coloring" of the graph using the anomaly scores computed using the latest observation. 

For this function, the order of assessment of the different kubernetes components plays an important role. Due to the existence of aggregator components, that require other components to be first evaluated as for their anomaly status before being themselves evaluated, an hierarchical approach is needed. This means that components are evaluated following the Kubernetes Resource Model starting from the bottom and going upwards.

\paragraph{Update Models}

When calling the "update models" function, the machine learning models used for prediction of anomaly scores are re-trained. Not all types of components have a model for their type. During the function call, a global dataframe containing all the accumulated observations from the graph is loaded. Then, an unsupervised model is initiated and is used to label these observations as anomalous or not. In case that the dataset does not contain any anomalous data points, a few outliers will be automatically created. This procedure is done by selecting a small subset of the data points and modifying their values by a few standard deviations. Afterwards, a Supervised model is created and trained to detect the anomalies using the labeled dataset. The trained supervised model is then returned for making predictions on future observations.

\paragraph{Components with Models}

The components with models are the ones that need to pass from a machine learning model in order to evaluate their anomaly status. Their workflow begins by querying the Neo4j database in order to obtain the latest observation for them. A pre processing step is then applied, that aggregates information, if needed, from components further down the hierarchy. 

Afterwards, the global observations database is updated with the latest observations. Those will be used in order to retrain the models. A maximum of observations is kept for each component type by discarding the old ones. The next step consists of standardization and classification. Standardization consists of subtracting the mean and dividing by the standard deviation of the observations used to train the model. Subsequently, they are fed to a supervised model in order to obtain their anomaly score. Last, the graph is updated with the new anomaly scores and the process continues with the next component type.

\paragraph{Aggregator Components}
Aggregator components are the ones that do not have a machine learning model for their type but simply gather the scores from components connected to them and update their anomaly score using the mean of the values they gathered. For the needs of the workflow, this procedure is done with the use of a Cypher Query. This query gets the scores from components connected to specific edge types and updates the anomaly score of the component being evaluated.

\paragraph{Configurability and Adaptability}

The frequency in which the models and the graph get updated can be specified by the user. The more often these updates take place, the more accurate the anomaly reporting will be, but the need for computational resources will increase as well. In addition, the user can also choose the metrics that are taken into account for the evaluation of each component type as well as the amount of historic observations used for retraining the models. This means that this work can be adapted for different infrastructures. The component is agnostic as to the nature of what it monitors and works with the assumption that anomalies will be few and infrequent. This component could be used to monitor activities ranging from security threats to hardware resource management with minimal reconfiguration.

% synexeia apo edo
\subsection{Models}

A slew of combinations of unsupervised and supervised models was explored, all implemented using the Scikit-learn Python library \cite{padregosa11}. More precisely, for the unsupervised part of the workflow, three anomaly detection, techniques were explored: Isolation Forests, DBSCAN (Density Based Spatial Clustering of Applications with Noise) and OCSVM (One Class Support Vector Machines). For the supervised part of the workflow, the classification models used for experiments are: Logistic Regression, SVM (Support Vector Machines) and Decision Trees \cite{theodoridis2020, alpaydin2020introduction}. All models were implemented using the Scikit-learn Python library \cite{sklearn_api}.

 \subsubsection{Unsupervised Models}

 For the unsupervised portion of the workflow, three anomaly detection, techniques were explored: 

 \begin{enumerate}
     \item \textbf{Isolation Forest:} Isolation forests are an unsupervised machine learning technique used for anomaly detection. They construct a collection of isolation trees and compute the average path lengths for each data point from these trees.  The trees are constructed by randomly partitioning the feature space. This ensures that points that are isolated from the rest will be partitioned faster as compared to normality points \cite{Liu:2008}.
        
     \item \textbf{DBSCAN (Density Based Spatial Clustering of Applications with Noise):} DBSCAN is an unsupervised machine learning technique primarily focused on clustering but it can also be used for anomaly detection problems. It groups together data points that form densities of arbitrary shapes and marks them as a cluster. Points that are isolated,  are not assigned to any clusters and are marked as outliers \cite{10.5555/3001460.3001507}.

    \item \textbf{OCSVM (One Class Support Vector Machines):} OCSVM are an unsupervised machine learning technique used for anomaly detection. They try to create a decision boundary that catches the majority of the points. They assume that anomalous points will not be placed inside the decision boundary, thus being marked as anomalies \cite{Bern:1999}.
    
 \end{enumerate}

 \subsubsection{Supervised Models}

For the supervised part of the workflow, three supervised classification models were used for experiments. These include:

 \begin{enumerate}
     \item \textbf{Logistic Regression:} Logistic Regression is a supervised machine learning technique used for binary classification tasks. It computes the weights of a linear function based on the training data. Then it uses this function to predict the binary outcome on future observations \cite{jurafsky-martin-2023}.
        
     \item \textbf{SVM (Support Vector Machines):}  SVM are a supervised machine learning algorithm that are used for classification. They find the hyperplane that maximizes the distance to the decision boundary between the classes of data points in the feature space  and use it to perform classification tasks on future observations \cite{liang-2016-svm-lecture, liang-2016-svm-lecture2}.
    
     \item \textbf{Decision Trees:} Decision trees are a supervised machine learning algorithm. They create a Directed acyclic graph where the non leaf nodes represent the input features, whose edges denote feature value ranges. Leaf nodes on the other hand belong to the output, hypothesis, space. New observations follow the path according to their features down the tree and the leaf node they reach constitutes the decision for that observation. Decision trees are well known for their interpretable nature \cite{Izza:2020}.
    
 \end{enumerate}

\section{Performance evaluation}
\begin{figure}[h!]
    \centering
    \includegraphics[width=0.6\linewidth]{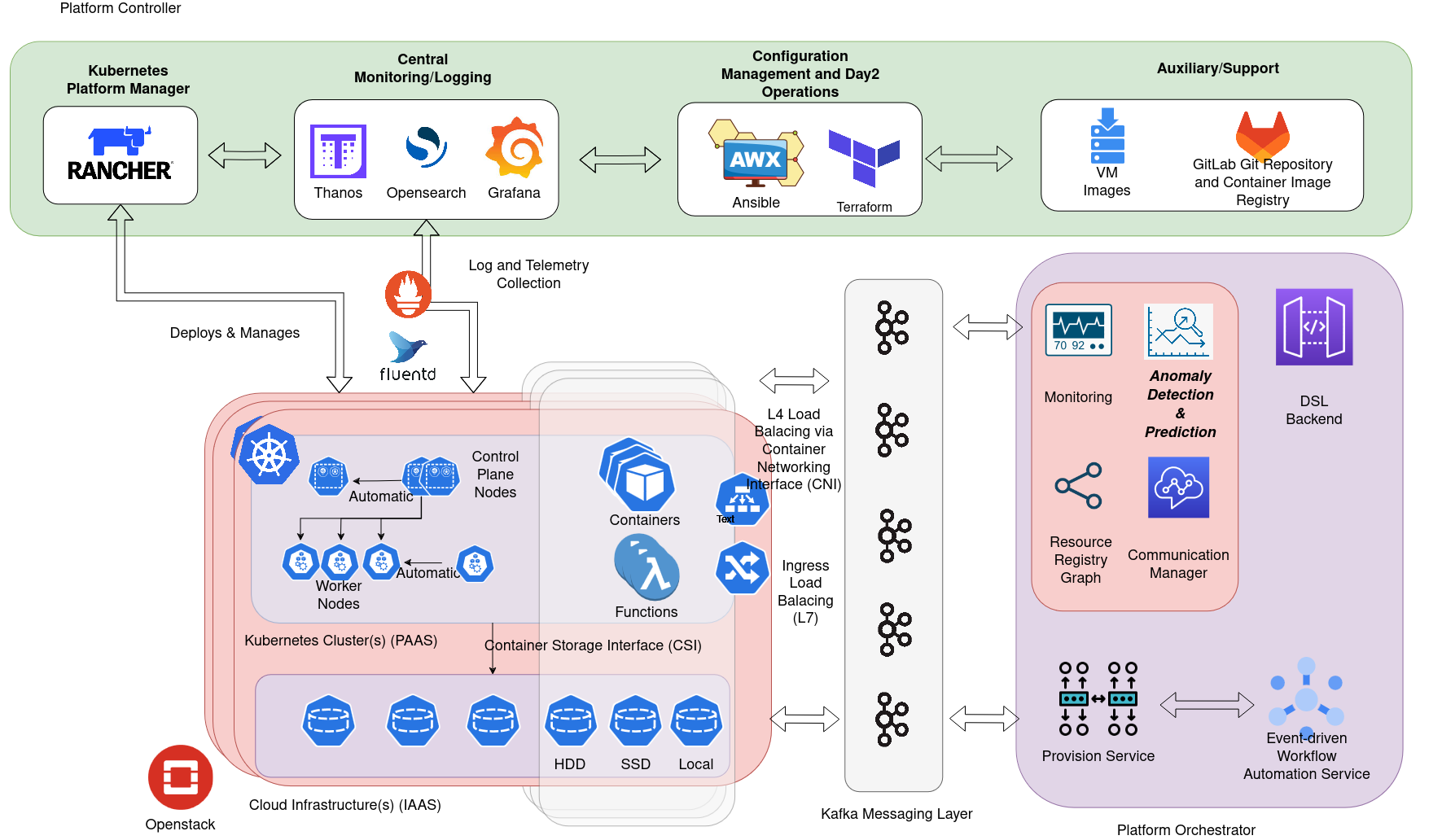 }
    \caption{Position of the Anomaly Detection and Prediction Component in the Architecture}
    \label{fig_arch}
\end{figure}

The proposed framewrok was developed and tested as part of EO4EU project. The (EO4EU) project is targeting a one-stop-shop approach to facilitate the discovery and processing of EO data. Consolidating massive amounts of data from EO data providers, its machine learning models dynamically extract patterns and insights that are easily accessible to users via improved user interfaces and extended reality. The processing needs of EO big data were based on a highly dynamic Kubernetes cluster with dynamic allocation of resources on the user needs. The monitoring, the resource registry, the anomaly detection and the graph representation were part of the architecture as shown in Fig. \ref{fig_arch}.

\subsection{Model Selection}

\subsubsection{Model Selection Experiment definition}

% maybe add that the EO4EU cluster was used to gather the data?

For the conduct of the experiments, metrics from a production cluster were taken. Each model was first given the chance to optimize its hyper parameters for the given nature of the data, before being compared to the other models. This was done using the hyper parameter optimization framework Optuna \cite{akiba:2019}. Each model got 50 trials in order to explore its hyper parameter space for the optimal combination given the data. 

Unsupervised models were evaluated based on their silhouette score, while the supervised ones were evaluated based on the F1 score. The first value is calculated using the mean distance of samples within the assigned class $a$ and the mean nearest class distance $b$ for the same samples. The silhouette coefficient is then computed using the formula 
\begin{equation}
   f= \frac{b-a}{\max(a,b)}
\end{equation}

$f$ provides a measure of how well clustering was performed with 1 signifying that all samples were appropriately placed and -1 indicating that the assignment was performed poorly. F1 score on the other hand, is a function of true and false predictions from the supervised model. It is computed as 
\begin{equation}
    F1 = \frac{\text{True Positive}}{\text{True Positive} + \frac{1}{2}(\text{False Positive} + \text{False Negative})}
\end{equation}

 The maximum possible F1 score is 1, computed when all samples are classified correctly. Both types of models were also evaluated in terms of the time required to be fitted and make predictions. 

In order to better understand the hyper parameter search process for each model, a contour plot is provided. Each axis corresponds to a hyper parameter, while dark areas on the plot signify better objective values. The contour lines represent regions of similar objective values. Denser lines mean that the different combinations give similar results. These plots give a visual way to understand how changes in one hyper parameter, effect the results of the objective value, in conjunction with another. 

\subsubsection{Unsupervised Models}

\begin{table}[ht]
\centering
\caption{Performance Metrics of Anomaly Detection Models}
\label{tab:anomaly_models}
\begin{tabular}{lcc}
\toprule
\textbf{Model}       & \textbf{Best Silhouette Value} & \textbf{Fit Time (s)} \\
\midrule
IsolationForest  & 0.7941             & 0.8232        \\
OneClassSVM      & 0.7302             & 0.1175        \\
DBSCAN           & 0.7661             & 0.0431        \\
\bottomrule
\end{tabular}
\end{table}

Table \ref{tab:anomaly_models}  describes the performance results for the Unsupervised models. All three models achieve similar silhouette values, with the Isolation Forest achieving the highest. 

These scores may appear relatively modest, compared to scores computed in a clustering task. This discrepancy could be attributed to the nature of the anomaly detection task. In anomaly detection the goal is to find outliers, in contrast to the task of clustering, which is to assign points to their nearest cluster center. Therefore, the given results could be considered as a meaningful achievement for the given task.

In terms of the time required to fit the models, the Isolation Forest model is by far the slowest of the three, with DBSCAN converging faster by a wide margin.

Since the model retraining is a rare occurrence in the Anomaly Detection and Prediction Workflow, as well as it does not require a small time window to be concluded, the Isolation Forest could be selected as the most suitable model for the task. The time penalty can be justified for the more accurate results.

 \begin{figure*}
   \centering
   \includegraphics[width=0.6\textwidth]{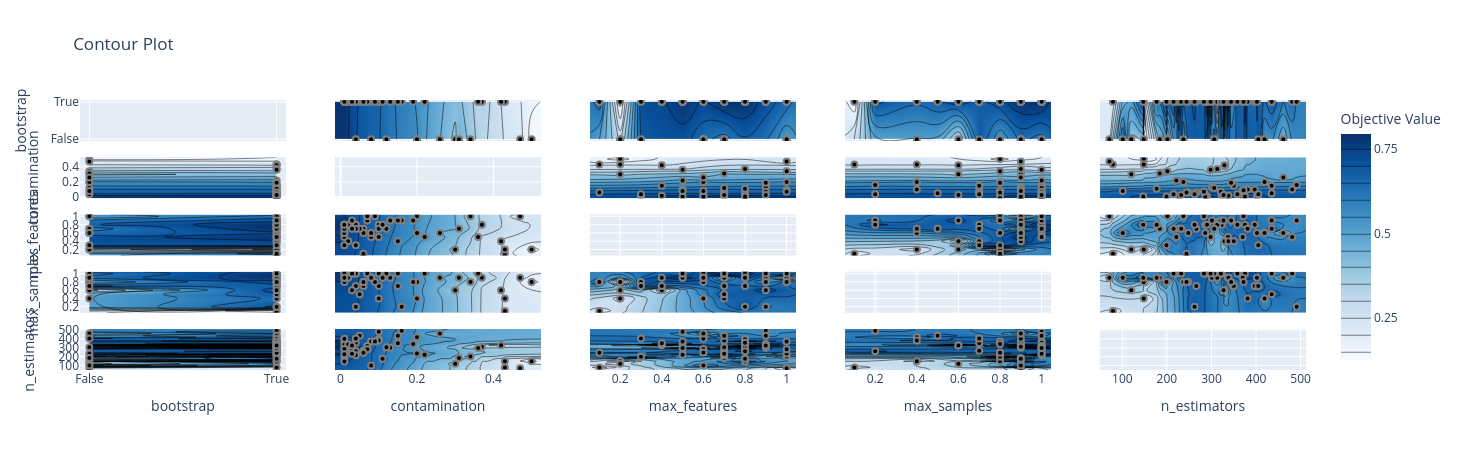}
   \caption{Contour plot for Isolation Forest}
   \label{fig:iforest}
 \end{figure*}

 \paragraph{Plot \ref{fig:iforest}} suggests that the Isolation Forest works best for the nature of the given task using: i) Bootstrap = "True", ii)Contamination = 0.01, iii) Max Features in $[0.6, 0.9]$, iv) Max samples in $[0.9, 1]$ and v) Estimators in $[250,350]$.

 The Contamination parameter has a big influence on the performance of the model, achieving the highest objective values while having its value remain stable at 0.1, no matter the combination with other hyper parameters. 

 \begin{figure*}
   \centering
   \includegraphics[width=0.6\textwidth]{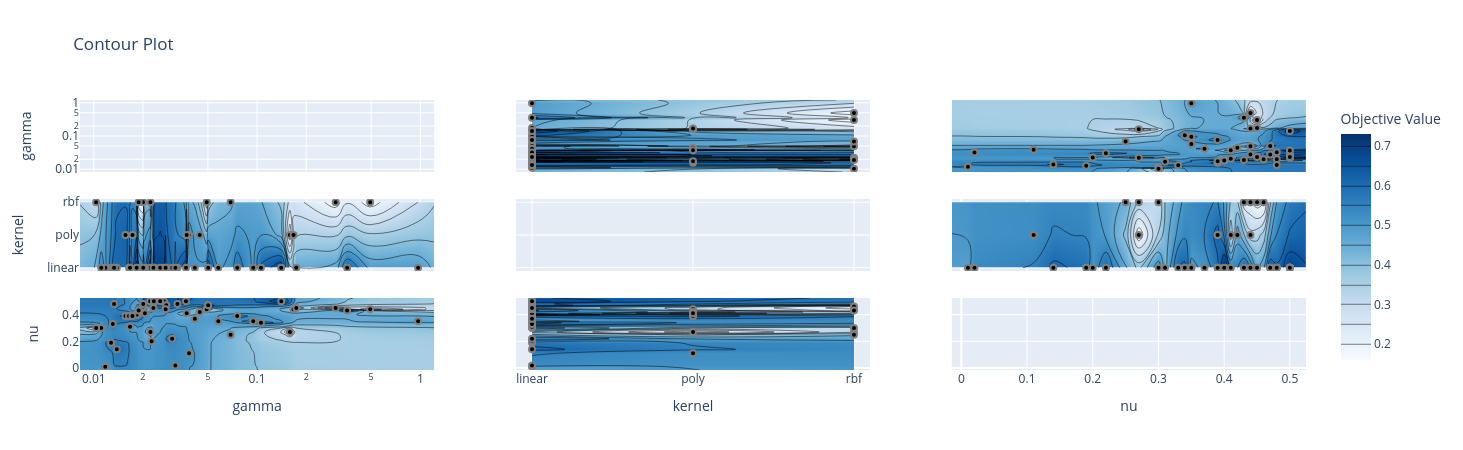}
   \caption{Contour plot for OneClassSVM}
   \label{fig:ocsvm}
 \end{figure*}

 \paragraph{Plot \ref{fig:ocsvm} } suggests that the One Class SVM model works best for the nature of the given task using: i) Bootstrap = "linear", ii) Gamma in $[0.01, 0.05]$ and nu in $[0.4,0.5]$.

 The kernel parameter has a big influence on the performance of the model, achieving the highest objective values while being set to "linear", with smaller influences from the other hyper parameters.

 \begin{figure*}
   \centering
   \includegraphics[width=0.6\textwidth]{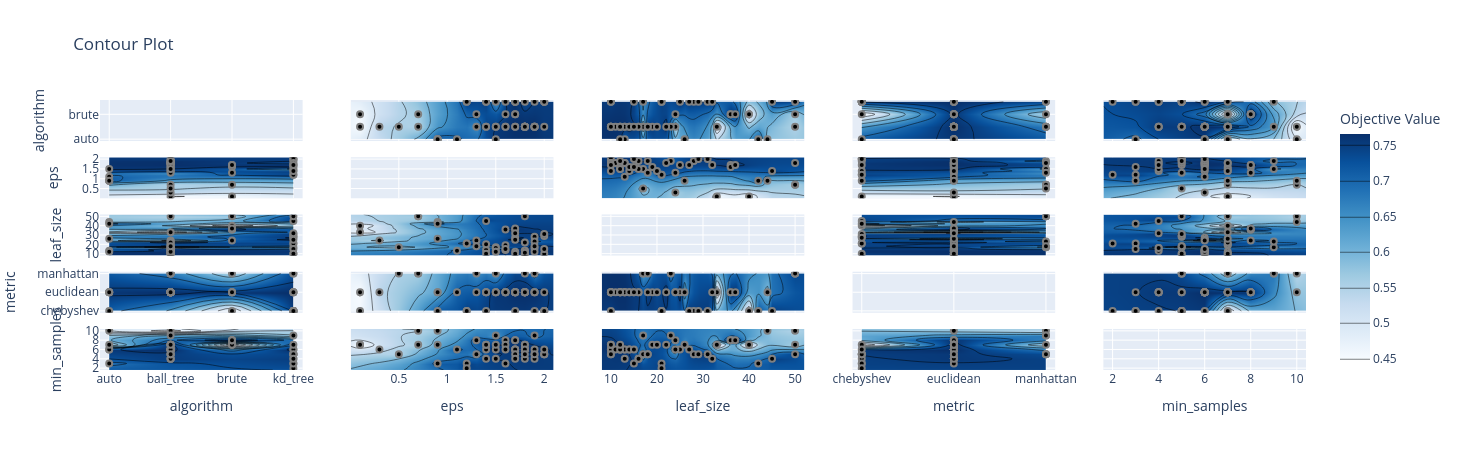}
   \caption{Contour plot for DBSCAN}
   \label{fig:dbscan}
 \end{figure*}

 \paragraph{Plot \ref{fig:dbscan}} suggests that DBSCAN works best for the nature of the given task using: i) Algorithm = "kd tree" or "ball tree", ii) eps in $[1.5,2]$, iii) Leaf size in $[10,15]$, iv) Metric = "euclidian" and v) min samples in $[0,8]$

 The eps parameter, meaning the maximum distance between to points for them to be considered neighbors, has a big influence on the performance of the model, achieving the highest objective values while having its value remain stable in the upper values range, no matter the combination with other hyper parameters. The metric and min samples parameters also assert some influence on the results, tho not as significant.

\subsubsection{Supervised Models}

\begin{table}[ht]
\centering
\caption{Performance Metrics of Classification Models}
\label{tab:classification_models}
\begin{tabular}{lccc}
\toprule
\textbf{Model}           & \textbf{Best F1 Score}  & \textbf{Fit Time (s)} & \textbf{Pred. Time (s)} \\
\midrule
Logistic Regress.  & 0.8702        & 0.0125          & 0.0038         \\
SVM                  & 0.8780        & 0.0565          & 0.0146         \\
Decision Tree        & 0.8864        & 0.0194          & 0.0025         \\
\bottomrule
\end{tabular}
\end{table}

Table \ref{tab:classification_models} showcases the results for the Supervised Model comparison. All three models achieved similar results in the classification tasks with the Decision Tree managing to outperform the other 2 models on the test set F1 score. The test set constitutes 20\% of the total dataset. 

In terms of the time for the fitting of the model with the training set, Logistic Regression comes on top, with a small margin over the Decision Tree. The fitting time is not as important for the same reasons as the time measured for the Unsupervised Models. 

The most important time, where being low is important, is the Prediction time. In terms of this time, Decision Trees come on top. This time is important due to the need to quickly update the graph with the new anomaly labels for each node. 

Taking into account all the performance metrics, the Decision Trees better fulfills the needs of this work, thus they will be selected for the anomaly prediction stage of the workflow. 

 \begin{figure*}
   \centering
   \includegraphics[width=0.8\textwidth]{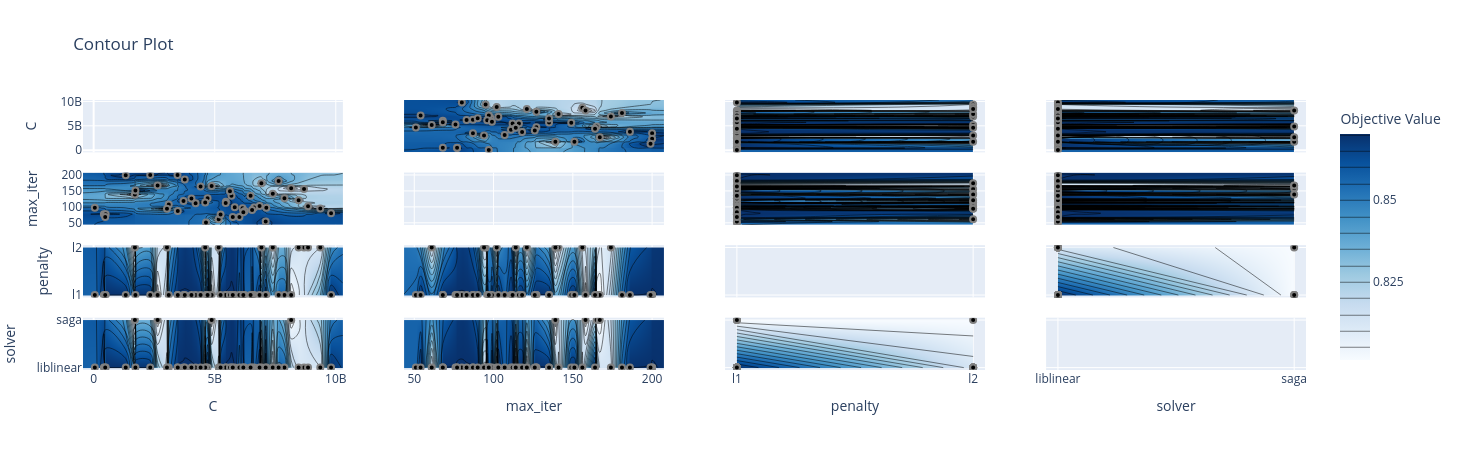}
   \caption{Contour plot for Linear Regression}
   \label{fig:supervised}
 \end{figure*}

 \paragraph{Plot \ref{fig:supervised}} suggests that the Logistic Regression model works best for the nature of the given task using: i) Penalty = "l1" and ii) Solver = "liblinear". 

 The solver and the penalty both had noticeable influence in the model's performance, with the solver skeweing the results the most. There were 2 more hyper parameters tested, C and max iterations but they did not have any impact in the model's performance. 

 \begin{figure*}
   \centering
   \includegraphics[width=0.8\textwidth]{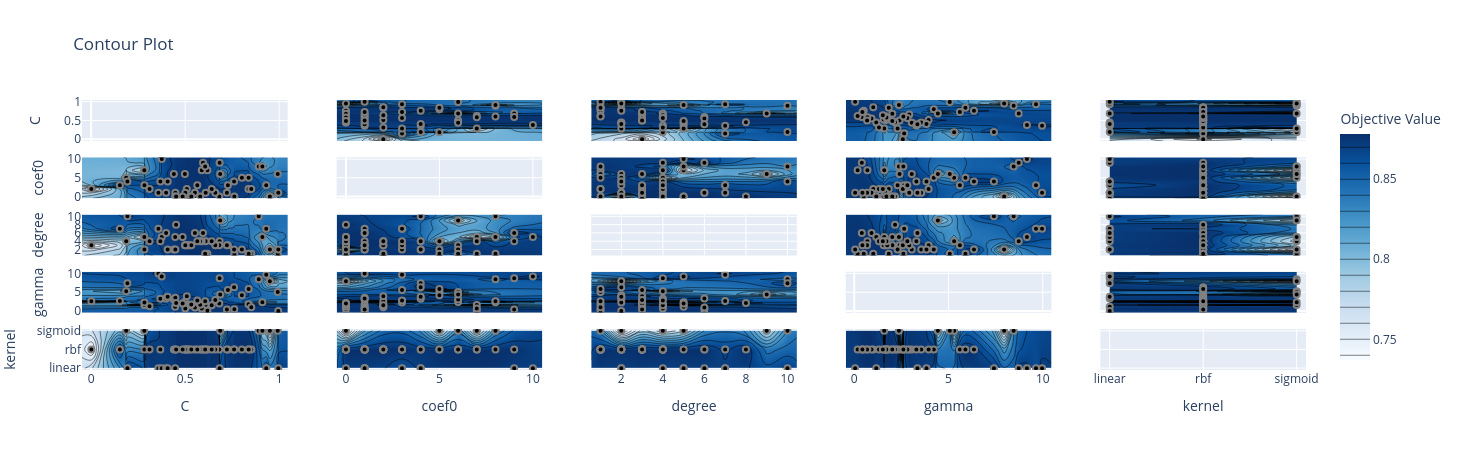}
   \caption{Contour plot for SVM}
   \label{fig:svm}
 \end{figure*}

 \paragraph{Fig. \ref{fig:svm}} suggests that the SVM model works best for the nature of the given task using: i) C in $[0.4,0.85]$ and ii) gamma in $[1,4.5]$ and iii) kernel = "rbf". 

 The kernel had the most significant impact in the model's performance in all the hyper parameter combinations. The C hyper parameter also had a noticeable impact in the model's performance. The degree and coef0 did not skew the results in any significant way.

 \begin{figure*}
   \centering
   \includegraphics[width=0.9\textwidth]{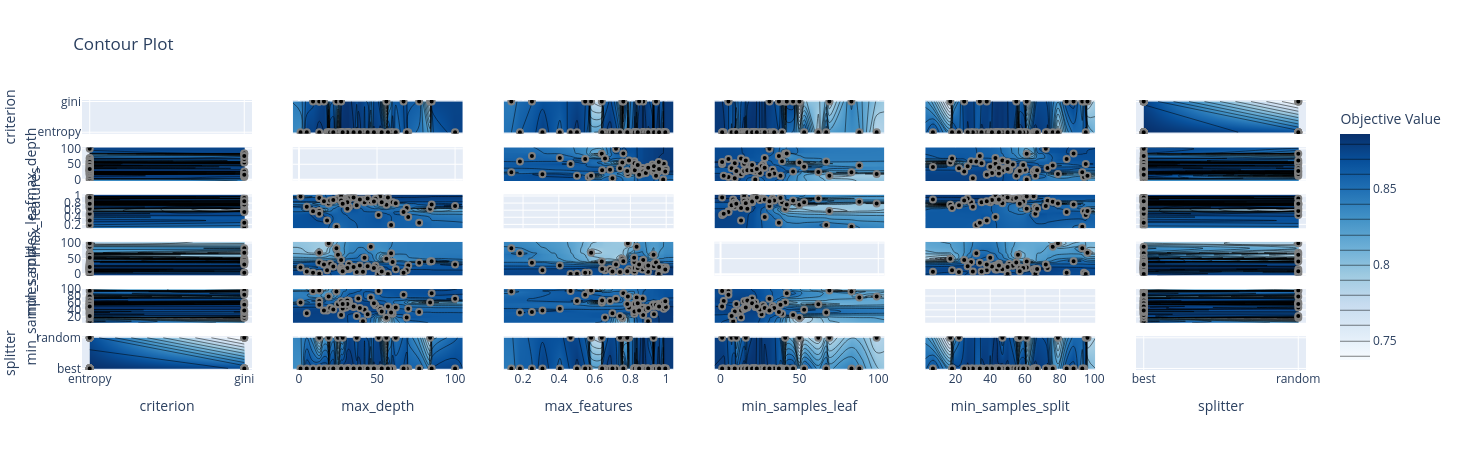}
   \caption{Contour plot for Decision Tree}
   \label{fig:tree}
 \end{figure*}

\paragraph{Fig. \ref{fig:tree}} suggests that the Decision Tree model works best for the nature of the given task using: i) Criterion = "entropy" and ii) splitter = "best", iii) max depth in $[5,35]$, iv) max features in $[0.6,1]$, v)min samples leaf in $[1,53]$ and vi) min samples split in $[20,80]$. 

The splitter hyper parameter had the biggest impact in the model's performance, followed by min samples leaf. The other hyper parameters did not have the same impact as the first 2 but were not insignificant.

\subsection{Experiments}
After the establishment of the individual benchmarking of the models, a combination of Decision Trees and Isolation Forest were applied in order to check the framework as end-to-end. The experiments described below were executed in EO4EU platform for $N=100$ and the average metrics are presented. Hence two types of experiments were conducted. For both experiments, a Pod was created that, at first, slept for 20 minutes and, then, used the stress command to simulate loads. For the first experiment, CPU load was simulated, by creating 32 Workers running the sqrt function. For the second experiment 32 workers did malloc calls and then hanged, making them unable to free the memory allocated. 

 \begin{figure}[!t]
 \centering
 \includegraphics[width=2.5in]{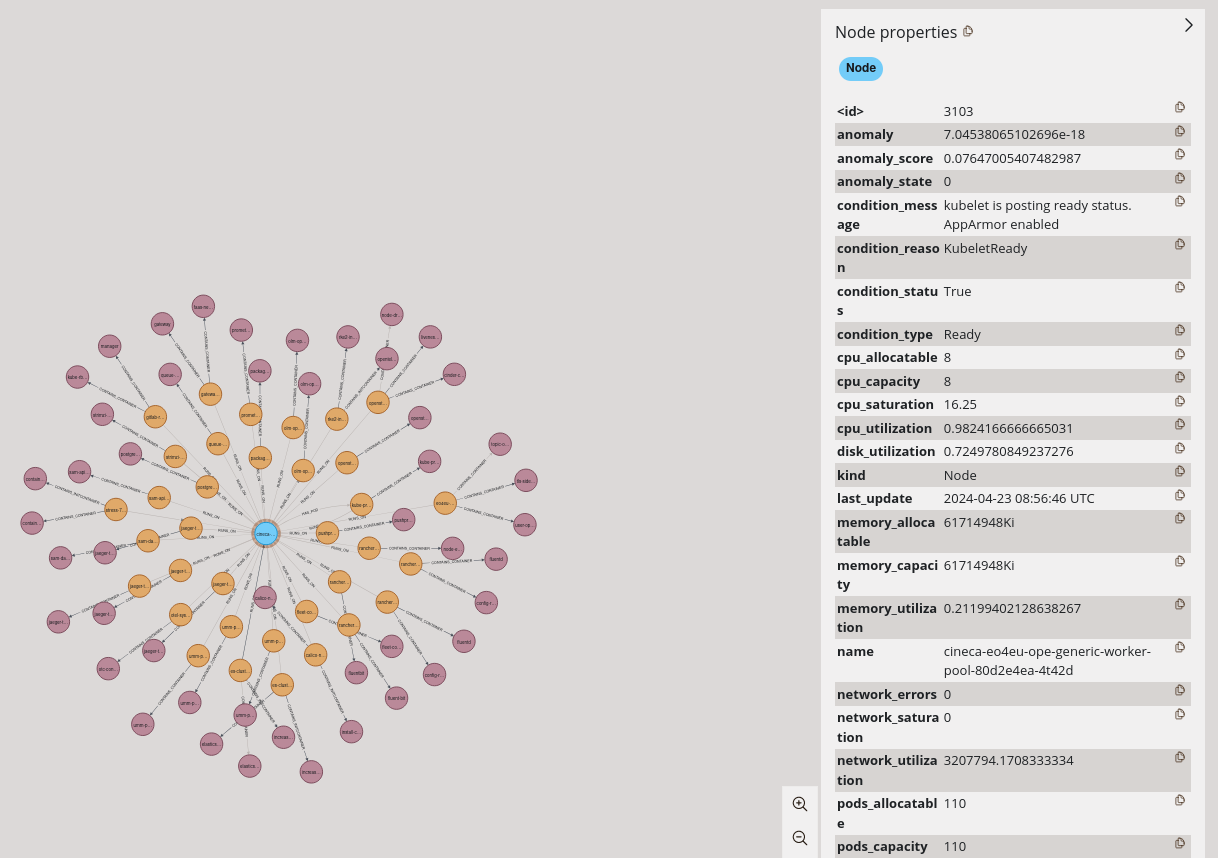}
 \caption{Node baseline before experiment starts}
 \label{baseline}
 \end{figure}

 \begin{figure}[!t]
 \centering
 \includegraphics[width=2.5in]{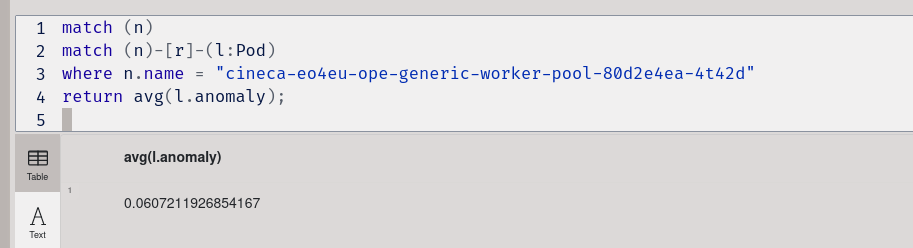}
 \caption{Average of anomaly scores of Pods in the targeted Node before experiments}
 \label{baseline_anom}
 \end{figure}

 \begin{figure}[!t]
 \centering
 \includegraphics[width=2.5in]{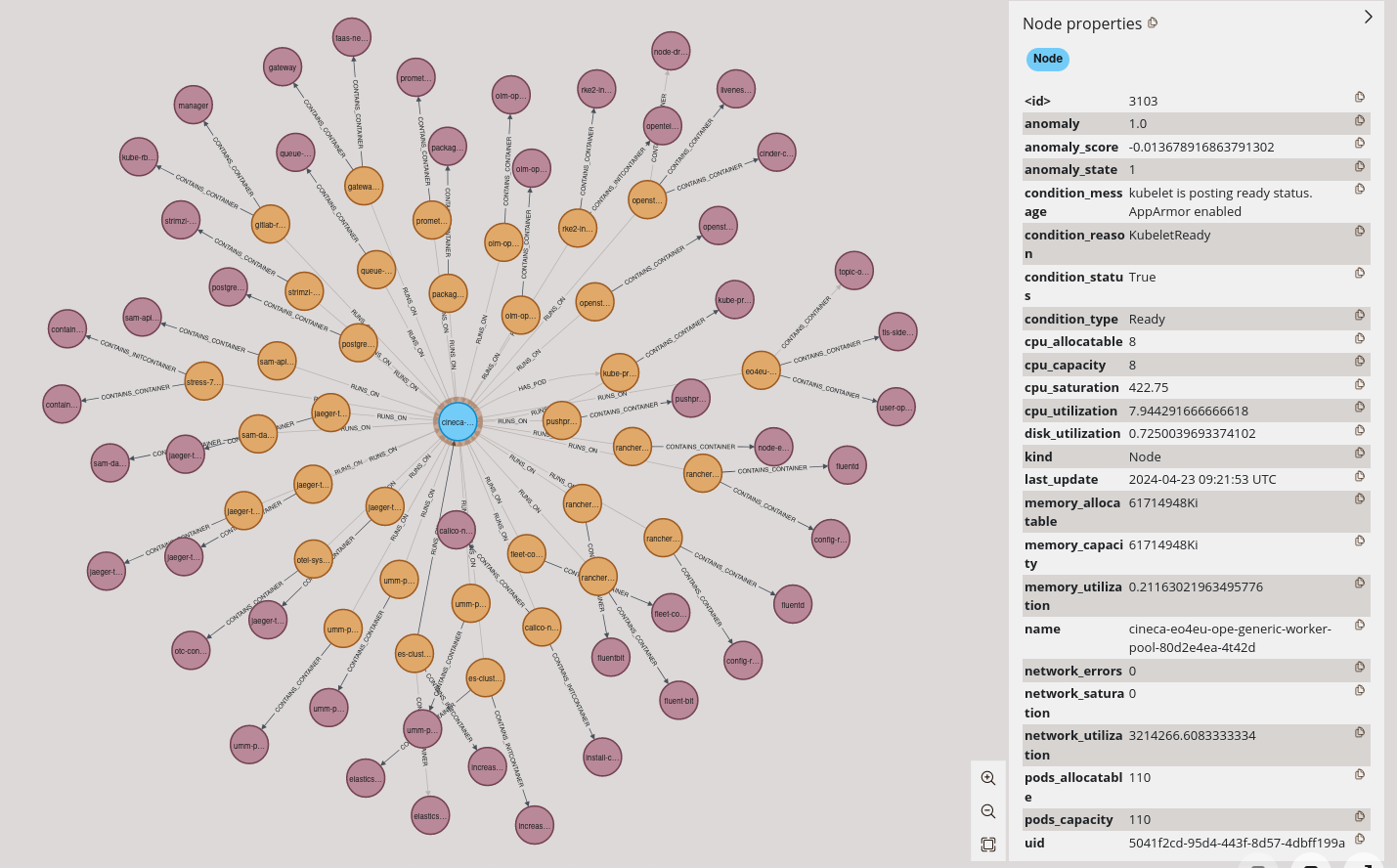}
 \caption{Node metrics under stress for first experiment.}
 \label{exp1_node}
 \end{figure}

 \begin{figure}[!t]
 \centering
 \includegraphics[width=2.5in]{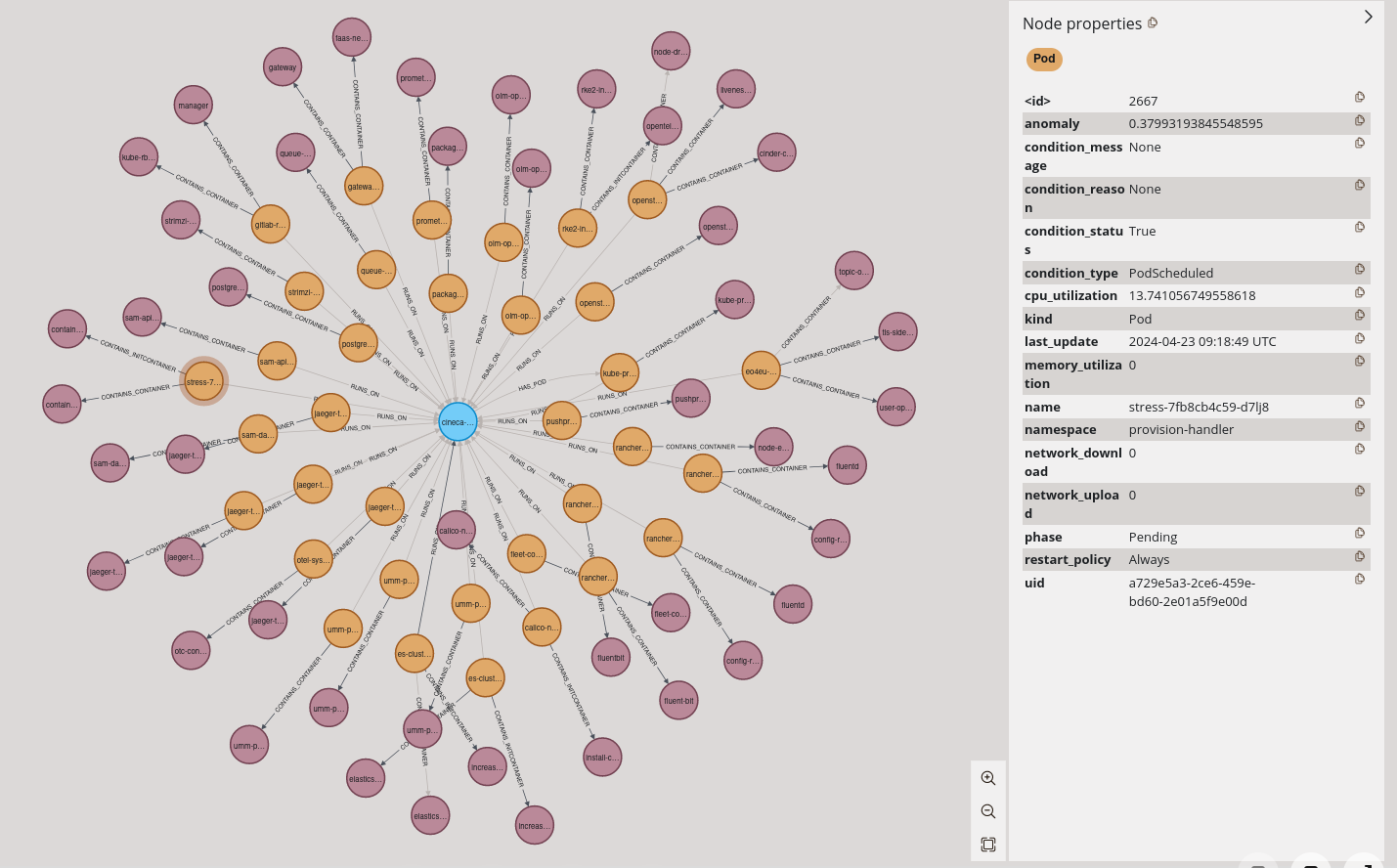}
 \caption{Pod metrics under stress for first experiment. Note, In the current version of the components, the $condition_{type}$ and phase information is not updated.}
 \label{exp1_pod}
 \end{figure}

 \begin{figure}[!t]
 \centering
 \includegraphics[width=2.5in]{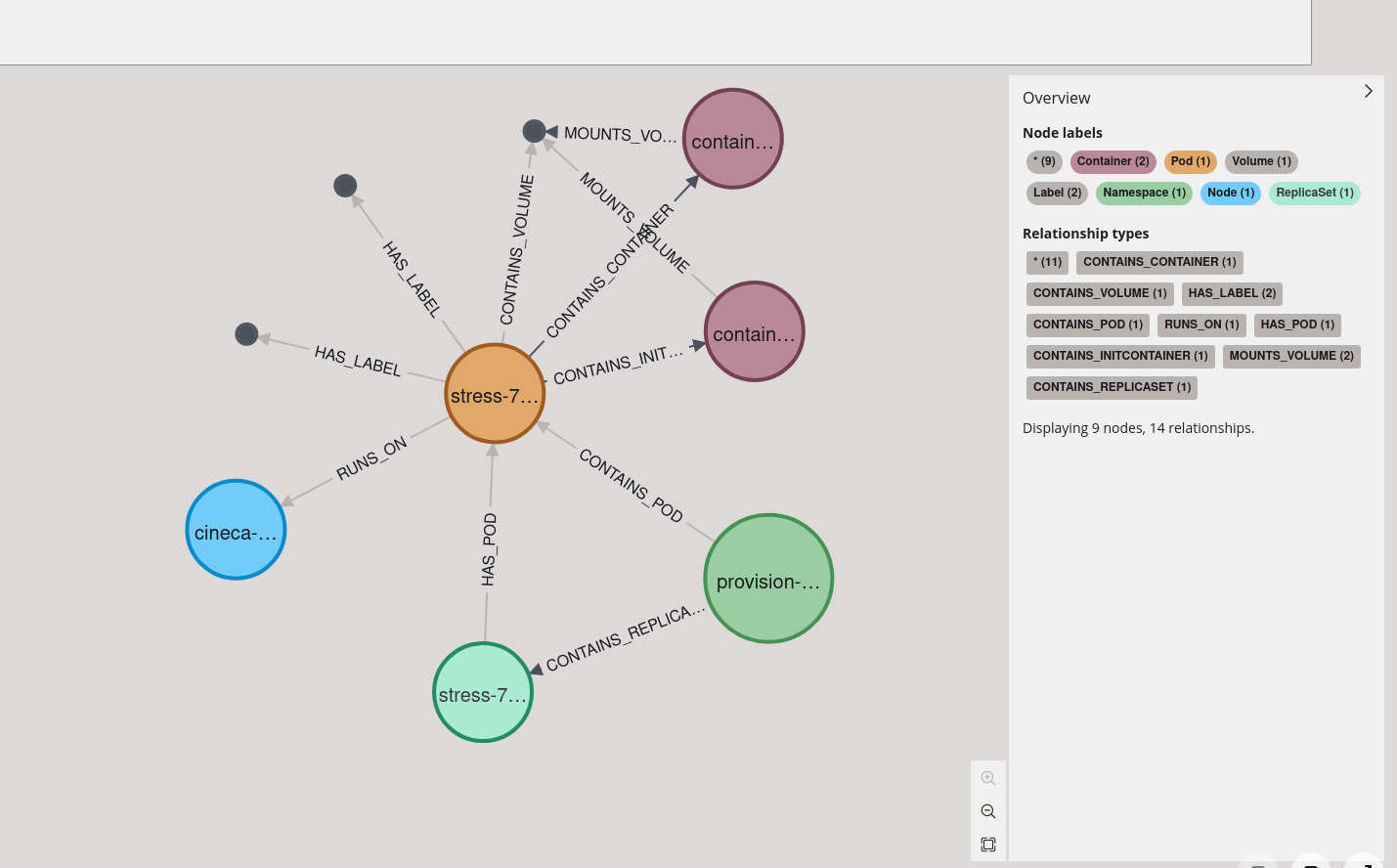}
 \caption{Graph Components connected to the problematic Pod found by looking at the edges of the Pod in the first experiment.}
 \label{family}
 \end{figure}

 \begin{figure}[!t]
 \centering
 \includegraphics[width=2.5in]{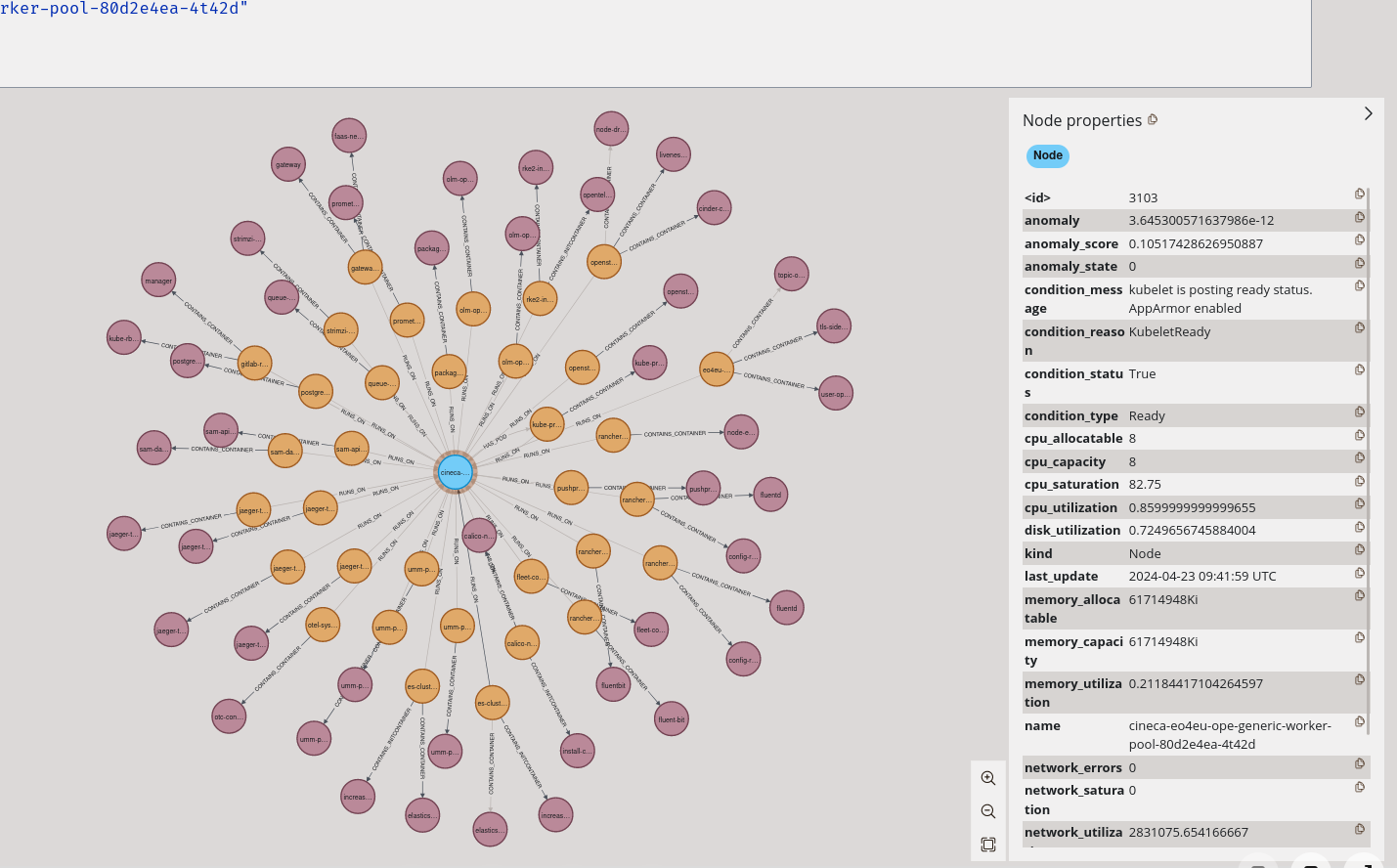}
 \caption{Node after the problematic Pod was removed in the first experiment.}
 \label{back_to_normal}
 \end{figure}

 \begin{figure}[!t]
 \centering
 \includegraphics[width=2.5in]{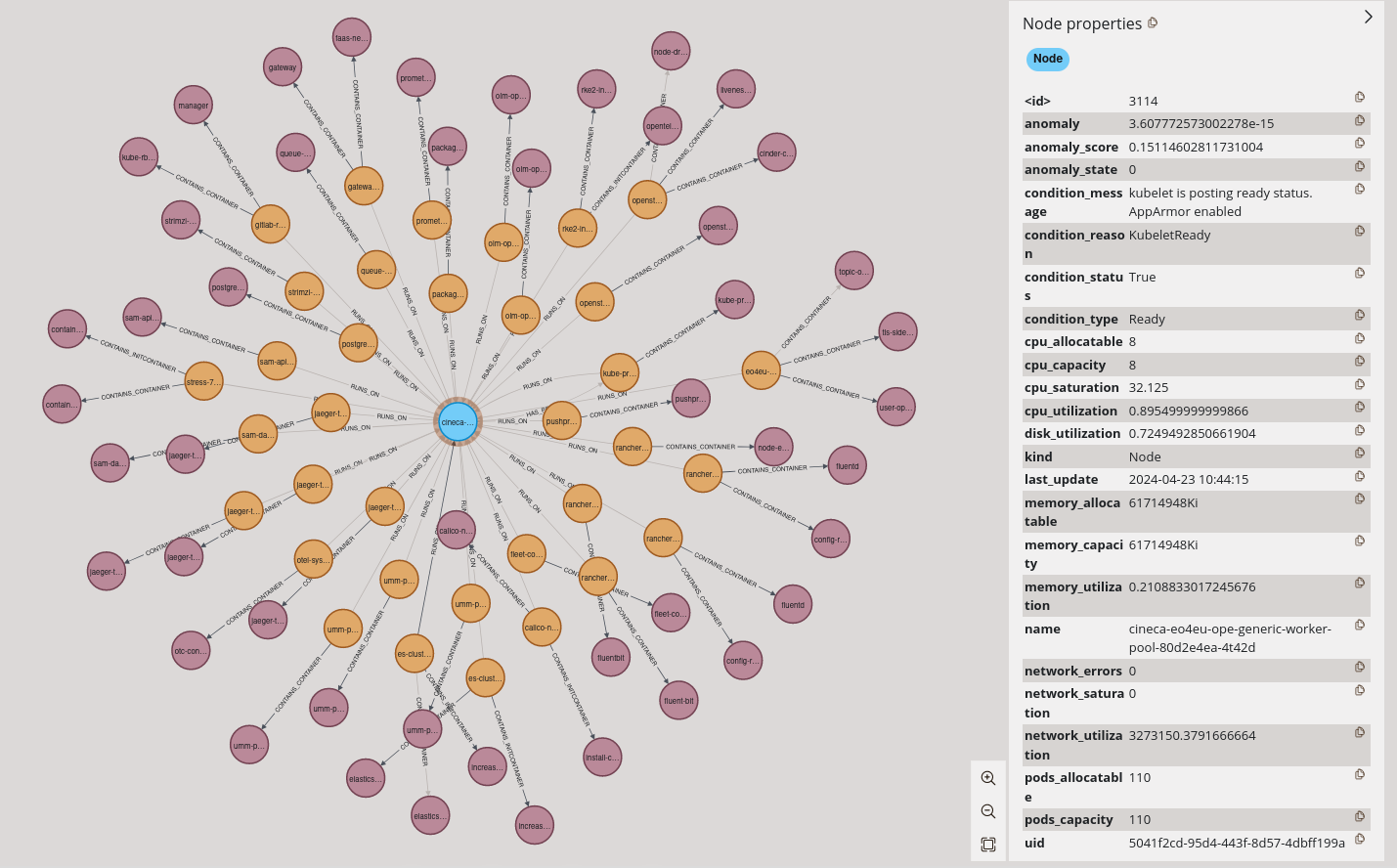}
 \caption{Node baseline in the second experiment.}
 \label{base2}
 \end{figure}

 \begin{figure}[!t]
 \centering
 \includegraphics[width=2.5in]{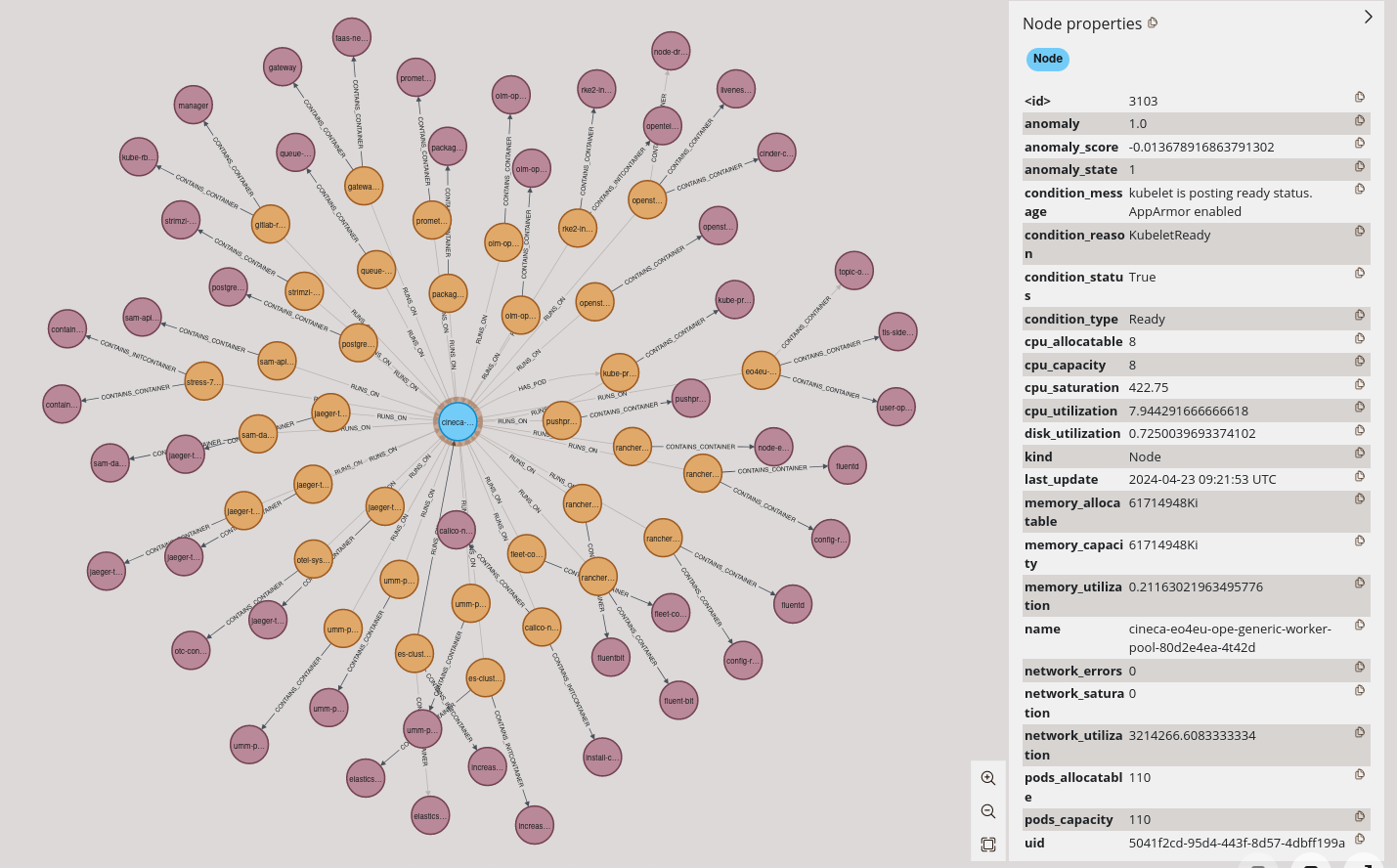}
 \caption{Node under stress in the second experiment.}
 \label{node2}
 \end{figure}

 \begin{figure}[!t]
 \centering
 \includegraphics[width=2.5in]{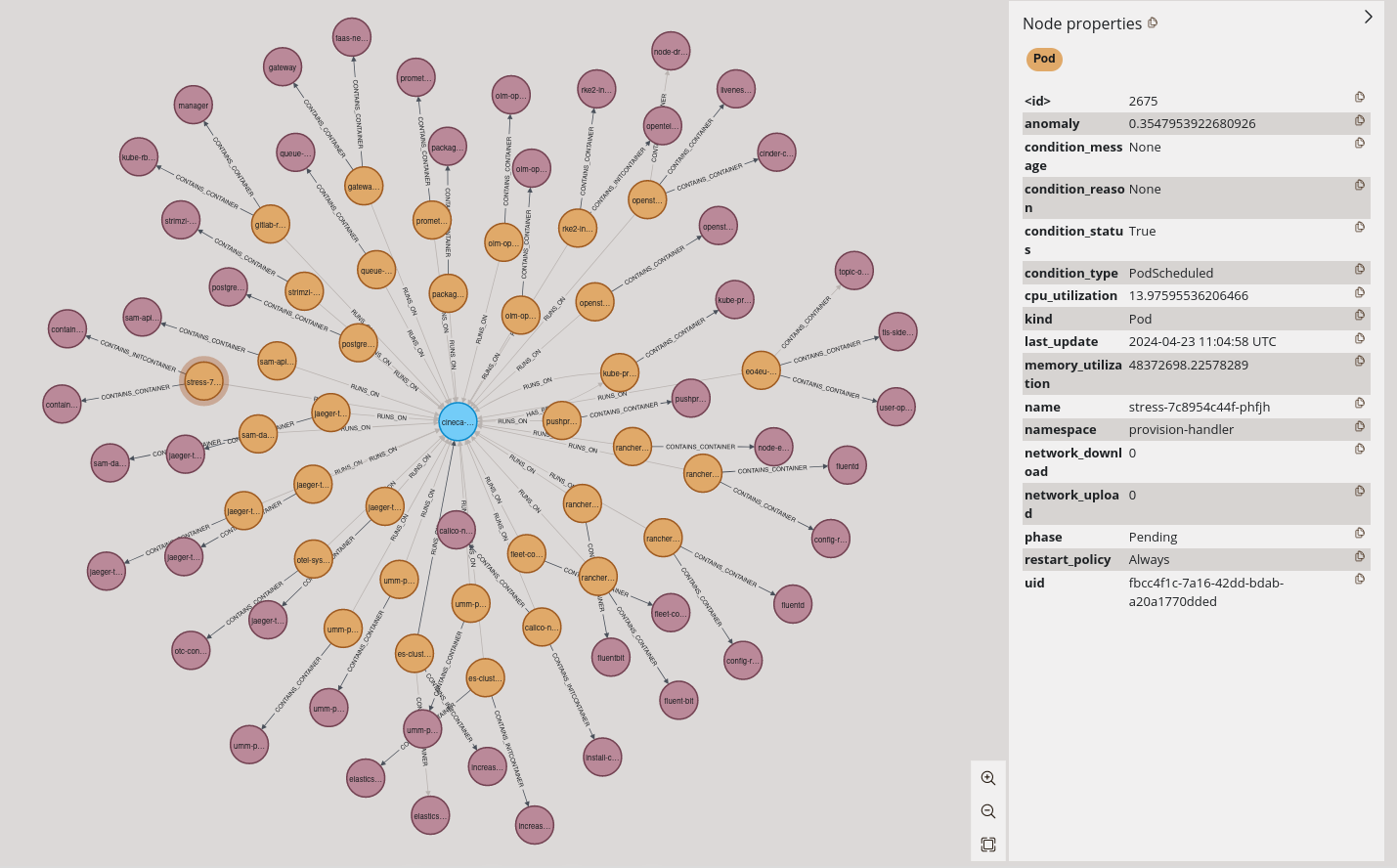}
 \caption{Pod under stress in the second experiment}
 \label{pod2}
 \end{figure}

Before running the experiments, a baseline of the Node's anomaly classification was established. The Node was deemed to be non anomalous with a high degree of confidence by the model.

\begin{figure}
    \centering
    \begin{subfigure}[t]{0.2\textwidth}
        \centering
        \includegraphics[width=1.5in]{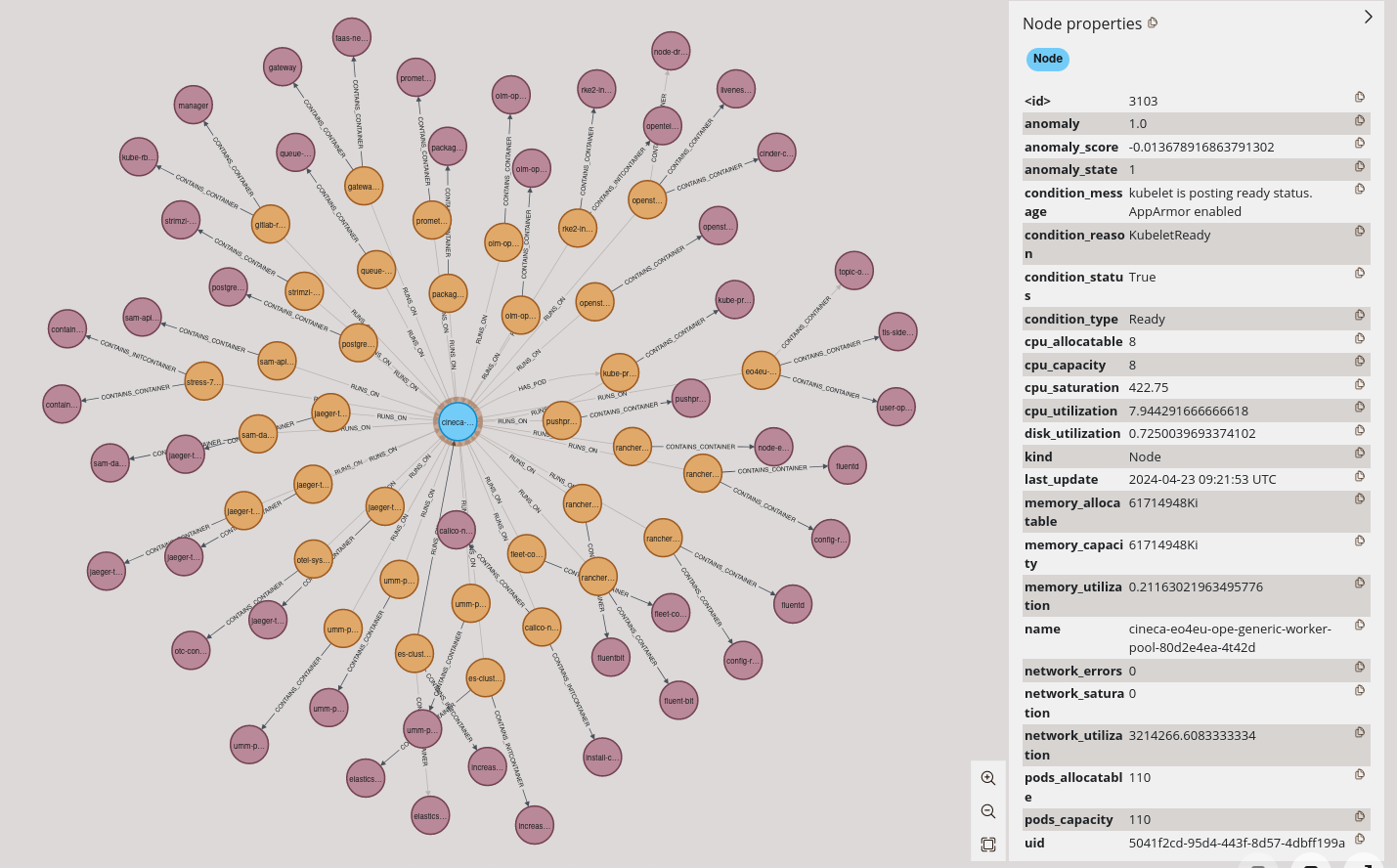}
        \caption{Node under stress in the second experiment.}
    \end{subfigure}
    ~
    \begin{subfigure}[t]{0.2\textwidth}
        \centering
        \includegraphics[width=1.5in]{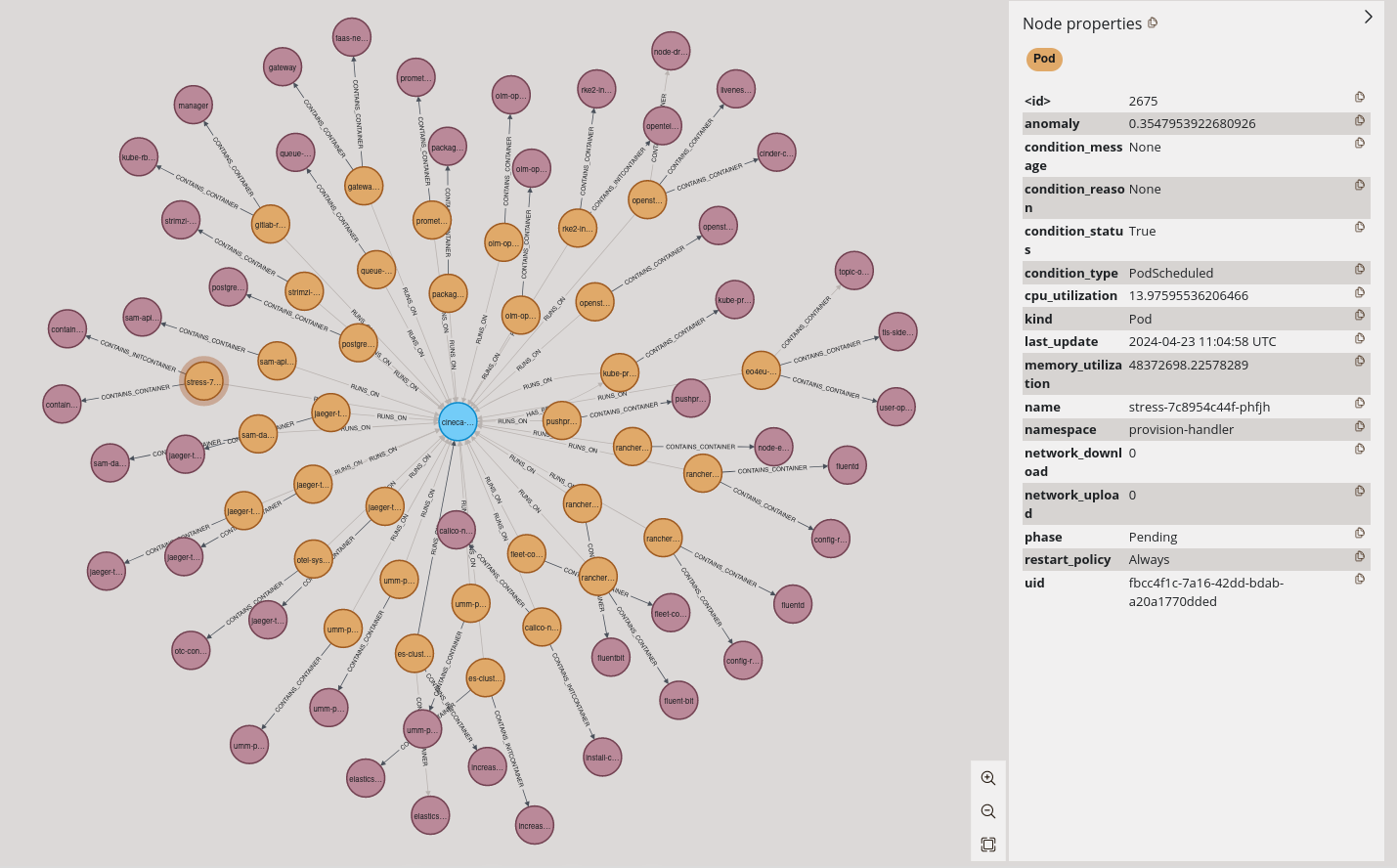}
        \caption{Pod under stress in the second experiment}
    \end{subfigure}
    \caption{Experiment 2}
\end{figure}

% status was taken. Its anomaly status was basically 0, signifying that it was not considered anomalous. Pods allocated to that Node had an average anomaly score of 0.06.

In the first experiment the Node was indicated as a definite anomaly. As for the Pod that caused the disturbance, it was identified by having an anomaly score that surpassed by orders of magnitude the anomaly scores of the other Pods allocated inside the same Node. The experiment, did not only highlight the Pod that was causing issues, but also manages to quickly identify the Namespace and the Replica Set-Deployment that caused the issues. The removal of the problematic Pod, returned the Node to its original anomaly classification of non-anomalous.

In the second experiment the Node that was used to host the Pod running the malloc script, was immediately identified as a definite anomaly. In addiction to the Node being identified, the responsible Pod also exhibited a score that quickly identified it as problematic, way above its peers in the same Node. 

 Both experiments managed to showcase the effectiveness of the proposed component. In both cases, the Node experiencing difficulties was identified by the anomaly detection mechanisms. The component also managed to track down the Pod causing the issues as well as the Namespace and the Replica Set-Deployment it belonged. In all the cases, the user could easily identify the problematic components from the visual database of Neo4j. By identifying the problematic components the system maintainer can take action. They can also be on the lookout for other components, connected to the ones in question, as to problems that may arise as a consequence of the problematic ones.

\section{Conclusion}

In this paper, we proposed and tested an anomaly detection and prediction component for Kubernetes Clusters. The evaluation was based on a live Application Cluster in order to test on real conditions and data. We started the evaluation by conducting experiments on model selection for both types of models. Subsequently, we tested the efficiency of the proposed solution by creating problematic Pods in a Node of the Cluster. The component was able to identify in all the cases the problematic Node as well as the defective Pod. In addition, it managed to find the Namespace the Pod was Deployed and the Deployment-Replica Set that it belonged enabling the system maintainer to take the appropriate action. 

In the future, we aim to enhance the scope of the component, by enhancing the graph representation with the parental layers of Kubernetes, like the Virtual Machines and OpenStack layers. This would give a deeper insight in the inner workings of the Cluster and provide even more information on the problematic parts, as well as, allow for better resource management by controlling the location of the deployments, in an effort to maximize efficiency. We also aim to enhance the anomaly prediction component with LSTM (Long Short Term Memory) neural networks. By creating a time series of graph observations, it is possible to predict future values and take appropriate action at an even earlier stage than it is possible now, with the current approach.

\section{Acknowledgments}
This work was supported by the EO4EU (AI-augmented ecosystem for Earth Observation data accessibility with Extended reality User Interfaces forService and data exploitation) project, funded by the EU’s Horizon Europe Research and Innovation Programme under Grant Agreement No. 101060784.

\bibliographystyle{elsarticle-num} 



\bibliography{main}
\end{document}